\documentclass[preprint]{revtex4}
\usepackage{graphicx}
\usepackage{ulem,color}
\usepackage{enumitem}
\usepackage{siunitx}
\usepackage{physics}
\usepackage{hyperref}

\usepackage{pgfplots} 

\allowdisplaybreaks

\newcommand{\e}{\mathrm{e}}
\newcommand{\coloneqq}{:=}

\newcommand{\R}{\text{1}}
\newcommand{\I}{\text{2}}

\newcommand{\T}{\mathcal{V}}

\begin{document}

\title{A tight Cram\'er-Rao bound for joint parameter estimation with a
  pure two-mode squeezed probe}

\author{Mark Bradshaw, Syed M. Assad, Ping Koy Lam}
\affiliation{Centre for Quantum Computation and Communication Technology, Department of Quantum Science,\\ Research School of Physics and Engineering, Australian National University, Canberra ACT 2601, Australia.}

\begin{abstract}
  We calculate the Holevo Cram\'er-Rao bound for estimation of the
  displacement experienced by one mode of an two-mode squeezed vacuum
  state with squeezing $r$ and find that it is equal to
  $4\exp(-2r)$. This equals the sum of the mean squared error obtained from a dual homodyne
  measurement, indicating that the bound is tight and that the
  dual homodyne measurement is optimal.
  
\end{abstract}

\maketitle

\section{Introduction}


In quantum mechanics, there is a limit to the precision to which we can simultaneously measure two observables that do not commute. If the observables are complementary variables, such as position and momentum, this limit is described by the Heisenberg uncertainty principle.

In continuous variable quantum optics, the bosonic quadrature operators $Q$ and $P$ are another pair of complementary variables, and obey the canonical commutation relation $[Q, P]=2i$, where throughout the paper we use units where $\hbar=2$. The bosonic field is also described by the annihilation operator $a$ and creation operator $a^\dagger$, which are related to the quadrature operators by $Q=a+a^\dagger$ and $P=i(a^\dagger-a)$. The displacement operator is given by
\begin{equation}
D(\alpha)=\exp(\alpha a^\dagger - \alpha^* a),
\end{equation}
where $\alpha=(q+ip)/2$ is the complex amplitude. 

In this paper we address this question: Given the probe state $\rho_0$ which undergoes a displacement of $D(\alpha)$
resulting in $\rho_\theta=D(\alpha)\rho_0 D^\dagger(\alpha)$, how well can we estimate the two parameters
$\theta_1\coloneqq q=2 \Re(\alpha)$ and $\theta_2 \coloneqq p=2 \Im(\alpha)$? Our figure of
merit is sum of the mean square error (MSE),
$\T \coloneqq E[(\hat\theta_1 -\theta_1)^2] + E[(\hat\theta_2-\theta_2)^2]$ where $E$ is the expectation value, and
$\hat \theta_\R$ and $\hat \theta_\I$ are the estimates of
$\theta_\R$ and $\theta_\I$ respectively. We aim to find a lower bound
to $\T$. These bounds are called Cram\'er-Rao bounds (CR bounds) and will only
depend on the state $\rho_\theta$ and independent of the measurement
performed on it. We calculate the CR bound based on the work of Holevo~\cite{Holevo1976a,Holevo2011}, which we call Holevo CR bound. First, we calculate the Holevo CR bound when the probe state $\rho_0$ is a single mode squeezed state, for which tight bounds are already known. Next, we calculate the Holevo CR bound when the probe is a two mode squeezed state. We find that it is superior to bounds calculated by previous authors~\cite{Genoni2013,Gao2014}; and that the bound can be reached by a simple measurement.

Our paper is divided up into sections as follows. In section \ref{sec_optics}, we briefly describe the Gaussian quantum optics used in our results. In section \ref{sec_estimation}, we summarize parameter estimation theory including CR bounds. In section \ref{sec_twomode}, we summarize the bounds found by other authors and the MSE from a dual homodyne measurement. In section \ref{sec_results}, we calculate the Holevo CR bound for one- and two-mode squeezed probes and discuss our results.

\section{Gaussian quantum optics}
\label{sec_optics}

Consider a state consisting of $m$ bosonic modes. Let the annihilation and creation operators of the $k$th mode be $a_k$ and $a^\dagger_k$, respectively, and the quadrature operators be $Q_k$ and $P_k$. Define a vector $Z$ to contain all the quadrature operators:
\begin{equation}
\vec{Z}=(Q_1,P_1,...,Q_m,P_m)
\end{equation}
The mean of the quadrature operators of a state $\rho$, otherwise known as the displacement vector of the state, is given by
\begin{equation}
M = \left[ \expval{Z_j} \right]_j,
\end{equation}
where $\expval{A}=\tr(\rho A)$ is the expectation value of operator $A$.
Define the covariance matrix $V$, which contains the variances of the quadrature operators, by
\begin{equation}
V = \left[\frac{1}{2}\expval{Z_j Z_k - Z_k Z_j} - \expval{Z_j}\expval{Z_k}\right]_{jk}.
\end{equation}

A thermal state is given by
\begin{equation}
\rho_\text{th}(N)=\frac{1}{N+1}\sum_{n=0}^{\infty}\left(\frac{N}{1+N}
\right)^n\ket{n}\bra{n}
\end{equation}
where $\ket{n}$ are the Fock states, and $N$ is the mean number of photons in the bosonic mode. A thermal state has a zero displacement vector and a covariance matrix of $V_\text{th}=(2N+1)I_2$ where $I_2$ is the $2\times2$ identity matrix.

The single-mode squeezing operator is given by
\begin{equation}
S(r)=\exp(\frac{1}{2}(r a^2-r {a^\dagger}^2)),
\end{equation}
where $r$ is the squeezing parameter. When acting on the vacuum state, this gives the squeezed vacuum state $\ket{S(r)} = S(r)\ket{0}$. The squeezed vacuum state has a zero displacement vector and a covariance matrix of
\begin{equation}
\label{eq_var_s}
V_{\text{sq}} =
\begin{pmatrix}
\e^{-2r} & 0 \\
0 & \e^{2r}
\end{pmatrix}.
\end{equation}

The two-mode squeezing operator is given by
\begin{equation}
S_2(r)=\exp(r a_1 a_2 -r a_1^\dagger a_2^\dagger).
\end{equation}
When acting on two vacuum states it gives the two-mode squeezed vacuum state, also known as the Einstein-Podolski-Rosen (EPR) state. The two-mode squeezed vacuum state has zero displacement vector and covariance matrix of
\begin{equation}
V_{\text{EPR}} =
\begin{pmatrix}
\cosh(2r) & 0 & \sinh(2r) & 0 \\
0 & \cosh(2r) & 0 & -\sinh(2r) \\
\sinh(2r) & 0 & \cosh(2r) & 0 \\
0 & -\sinh(2r) & 0 & \cosh(2r) 
\end{pmatrix}.
\end{equation}

A beam splitter is used to mix two modes. It is described by the unitary transformation
\begin{equation}
B(\phi)=\exp(\phi(a_1^\dagger a_2 - a_1 a_2^\dagger))
\end{equation}
where $\tau=\cos^2\phi$ is the transmissivity of the beam splitter.

\section{Parameter estimation theory}
\label{sec_estimation}

Let $\rho_\theta$ be a family of states parametrized by $d$ parameters $\theta=(\theta_1,\theta_2,...\theta_d)$. The goal of parameter estimation is to estimate the value of $\theta$ based on the outcome of a measurement on $\rho_\theta$. In quantum mechanics, a measurement is described by a positive operator-valued measure (POVM) $\Pi=\{\Pi_x\}$. Each measurement outcome $x$ has a corresponding non-negative hermitian operator $\Pi_x$ associated with it, where the probability of measuring $x$ on a state $\rho_\theta$ is $p_\theta(x)=\tr(\Pi_x\rho_\theta)$, and the POVM elements sum to Identity: $\sum_x \Pi_x=I$. We then need an estimator $\hat\theta(x)$, which maps the observed outcome $x$ to an estimate for $\theta$. An estimator is called locally unbiased at $\theta$ if $E[\hat\theta(x)]=\theta$ at the point $\theta$. An estimator is called unbiased if and only if it is locally unbiased at every $\theta$.

The MSE matrix $V_\theta[\hat{\theta}]$ of the estimator $\hat{\theta}$ is given by
\begin{equation}
V_\theta[\hat{\theta}] = \left[\sum_x p_\theta(x) (\hat{\theta_j}(x)-\theta_j)(\hat{\theta_k}(x)-\theta_k) \right]_{jk}.
\end{equation}
The sum of the MSE $\T$ is the trace of the MSE matrix:
\begin{equation}
\T=\Tr{V_\theta[\hat{\theta}]}.
\end{equation}
Here, $\Tr\left\{\cdot \right\}$ denotes the trace of an estimator matrix. The Cram\'er-Rao bound provides a lower bound to the MSE matrix for a classical probability distribution $p_\theta(x)$:
\begin{equation}
V_\theta[\hat\theta]\ge(J_\theta[p_\theta])^{-1}
\end{equation}
where $A\ge B$ for matrices $A$ and $B$ means $A-B$ is positive semi-definite. Taking the trace we get a bound for the MSE, 
\begin{equation}
\T\ge\Tr{(J_\theta[p_\theta])^{-1}}.  
\end{equation}
$J_\theta[p_\theta]$ is the classical Fisher information matrix is given by
\begin{equation}
J_\theta[p_\theta] = \left[\sum_x p_\theta(x) \frac{\partial \log p_\theta(x)}{\partial \theta_j} \frac{\partial \log p_\theta(x)}{\partial \theta_k}\right]_{jk}.
\end{equation}
This provides a bound to the MSE matrix for a fixed measurement $\Pi$. Next we define the most informative quantum Cram\'er-Rao bound, by minimizing over all POVMs.
\begin{equation}
C_\theta^{MI} = \min_{\Pi} \Tr\{J_\theta[\Pi]^{-1}\}
\end{equation}
In practice, this minimization is difficult to perform, so lower bounds are used instead. The first one is base of the symmetric log derivative (SLD) operator $L_{\theta,j}$ which is defined by
\begin{equation}
\frac{\partial \rho_\theta}{\partial \theta_j} = \frac{1}{2}\left( \rho_\theta L_{\theta,j} + L_{\theta,j} \rho_\theta \right).
\end{equation}
This is used to calculate the SLD quantum fisher information matrix defined by
\begin{equation}
G_\theta = [\expval{ L_{\theta,j},L_{\theta,k}}_{\rho_\theta}]_{jk},
\end{equation}
where we use an inner product defined by
\begin{equation}
\expval{X,Y}_{\rho_\theta} = \tr(\rho_\theta\frac{1}{2}(YX^\dagger+X^\dagger Y)),
\end{equation}
where $\tr(\cdot)$ denotes trace of a density matrix. This leads to a bound on the sum of the MSE, which we call the SLD CR bound $C_\theta^{S}$  \cite{Helstrom1967,Helstrom1969}.
\begin{equation}
\T\ge C_\theta^{S} = \Tr(G_\theta^{-1}).
\end{equation}
The next quantum CR bound we consider is based on the right log derivative (RLD) operator $\tilde{L}_{\theta,j}$ defined by
\begin{equation}
\frac{\partial \rho_\theta}{\partial \theta_j} = \rho_\theta \tilde{L}_{\theta,j}.
\end{equation}
This is used to calculate the RLD quantum fisher information matrix
\begin{equation}
\tilde{G}_\theta = [\langle \tilde{L}_{\theta,j}, \tilde{L}_{\theta,k}\rangle_{\rho_\theta}^+]_{jk},
\end{equation}
where we use an inner product defined by
\begin{equation}
\expval{X,Y}_{\rho_\theta}^+ = \tr(\rho_\theta YX^\dagger).
\end{equation}
This leads to a bound on the sum of the MSE \cite{Yuen1973}, which we call the RLD CR bound $C_\theta^R$, given by
\begin{equation}
\T \ge C_\theta^R = \Tr\{\Re \tilde{G}_\theta^{-1}\} + \textrm{TrAbs}\{\Im \tilde{G}_\theta^{-1}\},
\end{equation}
where $\text{TrAbs}\left\{\cdot \right\}$ denotes the sum of the absolute values of the eigenvalues of a matrix. 

While the RLD and SLD CR bounds are easy to compute~\cite{Paris2009,Petz2011}, in general they are not always achievable. The SLD CR bound corresponds to performing the optimal measurements for the estimation of each parameter ignoring the other. But for non-commuting observables, it might not be possible to perform the two optimal measurements simultaneously. Similarly, the RLD CR bound is in general not obtainable by a valid measurement. 

Holevo derived another bound for the MSE~\cite{Holevo1976a,Holevo2011}, which we shall call the Holevo CR bound. The Holevo CR bound is defined through the
following minimisation
\begin{align}
\label{eq_hol}
C_\theta^H \coloneqq \min_{\vec{X}\in \mathcal{X}} h_\theta[\vec{X}]
\end{align}
and  $\mathcal{X} \coloneqq \left\{(X_1,X_2,...X_d) \right\}$ where $X_j$ are Hermitian operators satisfying the locally unbiased conditions
\begin{align}
\label{eq_xcon1}
\tr(\rho_\theta X_j)&=0 \\
\label{eq_xcon2}
\tr(\frac{\partial \rho_\theta}{\partial\theta_j} X_k)&=\delta_{jk}\;
\end{align}
and $h_\theta$ is the function
\begin{align}
\label{eq_hol2}
h_\theta[\vec{X}] \coloneqq  \text{Tr}\left\{ \Re Z_\theta[\vec{X}]\right\} +\text{TrAbs} \left\{\Im Z_\theta[\vec{X}]\right\}\;.
\end{align}
$Z_\theta[\vec{X}]$ is a $d \times d$ matrix
\begin{align}
\label{eq_zmat}
Z_\theta[\vec{X}] \coloneqq \left[ \tr \left( \rho_\theta X_j X_k \right) \right]_{j,k}\;.
\end{align}
For any $\vec{X}$ satisfying the condition Eq.\ (\ref{eq_xcon2}), $h_\theta[\vec{X}]\ge C_\theta^S$ and $h_\theta[\vec{X}]\ge C_\theta^R$. At the minimum of $h_\theta$, defined above as $C_\theta^H$, $\T \geq C_\theta^H$. So, the Holevo CR bound is always greater than or equal to the RLD and SLD bounds. See for example \cite{Hayashi2006} for proof of the above statements.

The Holevo CR bound involves a minimization over the measurement
space, is in general hard to compute, and can be attained
by a collective measurement~\cite{Yamagata2013}. When the probe state
has rank one, an individual measurement is sufficient to attain the
Holevo CR bound~\cite{Matsumoto2002}.

\section{SLD and RLD CR bounds for two-mode squeezed probe}
\label{sec_twomode}

\begin{figure}
\centering
\newlength\figHeight 
\newlength\figWidth 
\setlength\figHeight{.4\linewidth} 
\setlength\figWidth{0.6\linewidth}
\input{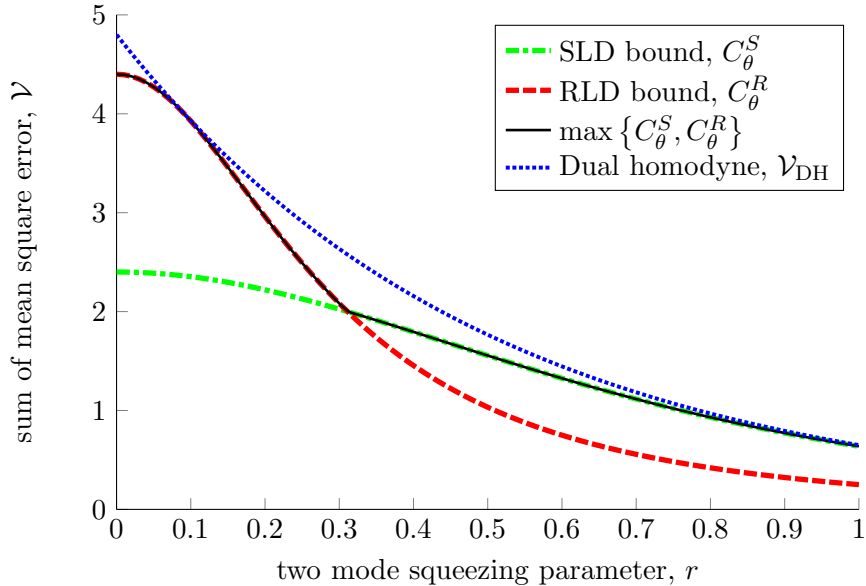}
\caption{Bounds for $\T$ as a function of squeezing parameter $r$ when
  the number of thermal photons $N=0.1$ are shown in green (SLD) and
  red (RLD). The black line is the most informative bound from these
  two bounds.  The sum of the mean squared error we get from a dual homodyne
  measurement is plotted in blue. We see that the bounds are not achieved by the dual homodyne measurement for some values of $r$. }
\label{tmb}
\end{figure}

Let the probe be a two-mode squeezed thermal state given by
$\rho_0=S_2(r) \left(\rho_\text{th}(N) \otimes \rho_\text{th}(N)
\right)S^\dagger_2(r)$, where if $N=0$ we get the two-mode squeezed vacuum. The first mode of the
probe state undergoes a unitary displacement operation $D(\theta)$ and ends up in the
state $\rho_\theta$. The SLD CR bound and RLD CR bounds are given by ~\cite{Genoni2013,Gao2014}:
\begin{align}
\label{conSt}C_\theta^S &\coloneqq \frac{2+4N}{ \cosh 2r}\\
\label{conRt}C_\theta^R &\coloneqq \frac{8N (1+N)}{(1+2N)
                                            \cosh 2r-1}.
\end{align}
Now let us consider a measurement which we call the dual homodyne measurement. The measurement consists of interfering the two modes on a beam splitter with transmissivity $\tau=1/2$, followed by homodyne measurement of the $Q$ quadrature of the first mode and a homodyne measurement of the $P$ quadrature of the second mode. The dual homodyne measurement gives $\T=(8N+4)\exp(-2r)\coloneqq\T_\text{DH}$. 

We plot $\T_\text{DH}$ and the two bounds in Fig.~\ref{tmb}. The dual homodyne measurement MSE does not reach the bounds for most values of $r$. This means that either the measurement is not optimal or the bounds are not tight. To help determine which is the case, we will calculate the Holevo CR bound and compare it to $\T_\text{DH}$.

\section{Results}
\label{sec_results}

Here, we calculate the Holevo CR bound for two cases: a pure single-mode squeezed probe and a pure two-mode squeezed probe.

\subsection{Calculation of Holevo bound for pure single-mode squeezed probe}
\label{sec_singlemode}
Consider a single mode squeezed probe
$\rho_0=S_1(r)\rho_\text{th}(N) S_1(r)^\dagger$. Applying the displacement operator $D(\theta)$, we
end up with the state $\rho_\theta=D(\theta)\rho_0 D^\dagger(\theta)$. For
pure states when $N=0$, this case is shown
in~\cite{Fujiwara1994a,Matsumoto2002,Fujiwara1999} to be {\it
  coherent} and the bound from the RLD is known to be
tight~\cite{Fujiwara1994a,Fujiwara1994,Fujiwara1999}. One optimal
joint estimation strategy is to alternatively apply the optimal
strategy for each parameter. In general, when $\rho$ is a mixed state,
the Cram\'er-Rao bounds are~\cite{Genoni2013,Gao2014}
\begin{align}
\label{conSt1}C_\theta^S &\coloneqq  (2+4N) \cosh 2r\\
\label{conRt1}C_\theta^R &\coloneqq  2+ (2+4N) \cosh 2r\;.
\end{align}
The RLD bound is always greater than the SLD bound and is hence a more informative lower bound. In fact, the dual homodyne measurement gives
$\T= 2+(2+4N)\cosh 2r$ which saturates the RLD bound~\cite{Genoni2013}. As the
bound increases with $r$, squeezed state probes perform worse than a
coherent state probe ($r=0$).

Although we already know what the result will be, as an exercise we
compute the Holevo bound for the pure single-mode squeezed
probe. In this case, $\rho_\theta=\ket{\psi}\bra{\psi}$ where $\ket{\psi}=D(\theta)\ket{S(r)}.$

The derivatives of the displacement operator $D(\theta)=\exp\left(\frac{i}{2} \theta_2 Q - \frac{i}{2} \theta_1 P \right) $ with respect to $\theta_1$ and $\theta_2$ are
\begin{align}
\frac{\partial}{\partial\theta_1} D(\theta) &= \left(-\frac{i}{2}P+\frac{i}{4}\theta_2\right) D(\theta) \\
\frac{\partial}{\partial\theta_2} D(\theta) &= \left(\frac{i}{2}Q-\frac{i}{4}\theta_1\right) D(\theta)  \;.
\end{align}
See appendix \ref{sec_dD} for the derivation. 

To simplify the calculation, we calculate the Holevo CR bound when $\theta$ is small, and hence evaluate at $\theta=0$, and assert that the bound will be the same for all $\theta$. The reason we can do this is because the Holevo CR bound is asymptotically attainable with an adaptive measurement scheme, given a set of $n$ identical states $\rho_\theta^{\otimes n}$ with $n\to\infty$. A rough estimate for $\theta$ can be obtained for a small number of measurements using $\sqrt{n}$ states, Then the remaining $n-\sqrt{n}$ states can be displaced by $D(-\tilde\theta)$ where $\tilde{\theta}$ is the rough estimate for $\theta$, resulting in states with a small $\theta$. 

We compute $\ket{\psi}$ and the derivatives of $\ket{\psi}$ with respect to $\theta_1$
and $\theta_2$ evaluated at $\theta=0$.
\begin{align}
\ket{\psi_0}&= \ket{\psi} \eval_{\theta=0} \nonumber \\
&= \ket{S(r)} \\
\ket{\psi_1}&= \frac{\partial}{\partial \theta_1} \ket{\psi} \eval_{\theta=0} \nonumber \\
&= -P  \ket{S(r)}\frac{i }{2}  \\
\ket{\psi_2}&= \frac{\partial}{\partial \theta_2} \ket{\psi} \eval_{\theta=0} \nonumber \\
&= Q  \ket{S(r)} \frac{i }{2}\;.
\end{align}
The inner products are
\begin{align}
\bra{\psi_0}\ket{\psi_0}&=1 \\
\bra{\psi_0}\ket{\psi_1}&=-\frac{i}{2} \bra{S(r)}P\ket{S(r)} \nonumber\\
&=0\\
\bra{\psi_0}\ket{\psi_2}&=\frac{i}{2} \bra{S(r)}Q\ket{S(r)} \nonumber\\
&=0\\
\bra{\psi_1}\ket{\psi_1}&=\frac{1}{4}\bra{S(r)}P^2\ket{S(r)} \nonumber\\
&=\frac{\e^{2r}}{4} \\
\bra{\psi_2}\ket{\psi_2}&=\frac{1}{4}\bra{S(r)}Q^2\ket{S(r)} \nonumber\\
&=\frac{\e^{-2r}}{4},
\end{align}
where we have used that the displacement vector of $\ket{S(r)}$ is zero, and the covariance matrix given by Eq.~(\ref{eq_var_s}).
\begin{equation}
\bra{\psi_1}\ket{\psi_2}=-\frac{1}{4}\bra{S(r)}PQ\ket{S(r)}.
\end{equation}
From the commutation relation $[Q,P]=2i$, we get $\Im  \expval{PQ}  = -1$.
The covariance of $Q$ and $P$ is given by
\begin{align}
V_{QP}&= \frac{1}{2} \expval{QP+PQ} - \expval{Q}\expval{P} \\
 &= \Re \expval{PQ}  - \expval{Q}\expval{P}\;.
\end{align}
From Eq.~(\ref{eq_var_s}) this should equal zero, and since the displacement vector is also zero, $\Re{\bra{S(r)}PQ\ket{S(r)}}=0$, so we have that
\begin{equation}
\bra{\psi_1}\ket{\psi_2}=\frac{i}{4}.
\end{equation}
We introduce a set of orthonormal vectors $\{\ket{e_0},\ket{e_1}\}$ such that
\begin{align}
\ket{\psi_0}&=\ket{e_0} \\
\ket{\psi_1}&=\ket{e_1}\frac{\e^{r}}{2} \\
\ket{\psi_2}&=\ket{e_1}\frac{i\e^{-r}}{2}\;,
\end{align}
which satisfies the inner products. With this, the constraint Eq.~(\ref{eq_xcon1}) becomes
\begin{align}
\bra{e_0}X_1\ket{e_0}&=0\\
\bra{e_0}X_2\ket{e_0}&=0,
\end{align}
The density matrix for $\ket\psi$ and its derivatives at the point $\theta=0$ are
\begin{align}
\rho_0&= \rho \eval_{\theta=0} \nonumber \\
&= \ket{\psi_0}\bra{\psi_0} \\
\rho_1&= \frac{\partial}{\partial \theta_1} \rho \eval_{\theta=0} \nonumber \\
&= \ket{\psi_0}\bra{\psi_1} + \ket{\psi_1}\bra{\psi_0} \\
\rho_2&= \frac{\partial}{\partial \theta_2} \rho \eval_{\theta=0} \nonumber \\
&= \ket{\psi_0}\bra{\psi_2} + \ket{\psi_2}\bra{\psi_0}\;.
\end{align}
The constraint Eq.~(\ref{eq_xcon2}) is therefore
\begin{align}
\bra{e_0}X_1\ket{e_1}&=\e^{-r}\\
\bra{e_0}X_2\ket{e_1}&=-i\e^{r}.
\end{align}
Because we are interested in the minimization of Eq.\ (\ref{eq_hol}), we
can set to zero all components of $X_1$ and $X_2$ not involved in the
constraints or not complex
conjugates of components involved in constraints. The minimization is trivial, and the solution occurs when
\begin{equation}
Z_\theta=\begin{pmatrix}
\e^{-2r} & i\\
-i&\e^{2r}
\end{pmatrix}\,.
\end{equation}
The Holevo CR bound for a pure single-mode squeezed state probe is therefore
\begin{equation}
C_\theta^H= 2+2 \cosh 2r\,,
\end{equation}
which equals the RLD bound and the variance from a dual homodyne
measurement when $N=0$ as expected.

\subsection{Calculation of Holevo CR bound for pure two-mode squeezed probe}
To calculate the Holevo CR bound for a pure two-mode squeezed probe, we follow a similar procedure as the single-mode case. The two-mode probe state can be transformed into a product state of
two single-mode squeezed probes by a beam splitter with trasmissivity $\frac{1}{2}$. The beam splitter is a unitary transformation, which does not affect the Holevo CR bound. Furthermore, when
$N=0$, $\rho$ has rank one, and this transformed version of $\rho$ can be
written as $U \rho U^\dagger = \ket{\psi}\bra{\psi}$ where
\begin{equation}
\ket{\psi}=D(\theta/\sqrt{2})\ket{S(r)}\otimes D(-\theta/\sqrt{2})\ket{S(-r)},
\end{equation}
We compute $\ket{\psi}$ and the derivatives of $\ket{\psi}$ with respect to $\theta_1$
and $\theta_2$ evaluated at $\theta=0$.
\begin{align}
\ket{\psi_0}&= \ket{\psi} \eval_{\theta=0} \nonumber \\
&= \ket{S(r)}\ket{S(-r)} \\
\ket{\psi_1}&= \frac{\partial}{\partial \theta_1} \ket{\psi} \eval_{\theta=0} \nonumber \\
&=-P \ket{S(r)}\ket{S(-r)}\frac{i}{2\sqrt{2}}  + \ket{S(r)}P\ket{S(-r)} \frac{i}{2\sqrt{2}} \\
\ket{\psi_2}&= \frac{\partial}{\partial \theta_2} \ket{\psi} \eval_{\theta=0} \nonumber \\
&=Q \ket{S(r)}\ket{S(-r)}\frac{i}{2\sqrt{2}}  - \ket{S(r)}Q\ket{S(-r)}  \frac{i}{2\sqrt{2}}\;.
\end{align}
Of interest are the inner products involving the states $\ket{\psi_0}$, $\ket{\psi_1}$ and $\ket{\psi_2}$, so we calculate them now.
\begin{align}
\bra{\psi_0}\ket{\psi_0}&=1 \nonumber \\
\bra{\psi_0}\ket{\psi_1}&=-\frac{i}{2\sqrt{2}}\bra{S(r)}P\ket{S(r)}+\frac{i}{2\sqrt{2}} \bra{S(-r)}P\ket{S(-r)} \\
&=0 \\
\bra{\psi_0}\ket{\psi_2}&=\frac{i}{2\sqrt{2}}\bra{S(r)}Q\ket{S(r)}-\frac{i}{2\sqrt{2}} \bra{S(-r)}Q\ket{S(-r)} \\
&=0 \\
\bra{\psi_1}\ket{\psi_1}&=\frac{1}{8}\bra{S(r)}P^2\ket{S(r)}+\frac{1}{8}\bra{S(-r)}P^2\ket{S(-r)} \nonumber \\
&=\frac{1}{8}\e^{2r}+\frac{1}{8}\e^{-2r} = \frac{\cosh 2r}{4} \\
\bra{\psi_2}\ket{\psi_2}&=\frac{1}{8}\bra{S(r)}Q^2\ket{S(r)}+\frac{1}{8}\bra{S(-r)}Q^2\ket{S(-r)} \nonumber \\
&=\frac{1}{8}\e^{-2r}+\frac{1}{8}\e^{2r} = \frac{\cosh 2r}{4} \\
\bra{\psi_1}\ket{\psi_2}&=-\frac{1}{8}\bra{S(r)}PQ\ket{S(r)}-\frac{1}{8}\bra{S(-r)}PQ\ket{S(-r)} \nonumber \\
&=\frac{i}{4}\;.
\end{align}
To satisfy the inner products, we introduce an orthonormal set of states $\{\ket{e_0},\ket{e_1},\ket{e_2}\}$ such that
\begin{align}
\ket{\psi_0}&=\ket{e_0} \nonumber \\
\ket{\psi_1}&=\ket{e_1}\frac{\cosh r}{2} + \ket{e_2}\frac{\sinh r}{2} \nonumber \\
\ket{\psi_2}&=\ket{e_1} \frac{i\cosh r}{2} -\ket{e_2} \frac{i\sinh r}{2} \label{eq_solc}
\end{align}
Using the construction in Eq.~(\ref{eq_solc}) the constraint Eq.~(\ref{eq_xcon1}) becomes
\begin{align}
\bra{e_0}X_1\ket{e_0}&=0 \label{eq_con1} \\
\bra{e_0}X_2\ket{e_0}&=0,
\end{align}
and the constraint Eq.~(\ref{eq_xcon2}) becomes
\begin{align}
\Re(\cosh r\bra{e_0}X_1\ket{e_1}+\sinh r \bra{e_0}X_1\ket{e_2})&=1 \label{eq_con3} \\
\Re(\cosh r\bra{e_0}X_2\ket{e_1}+\sinh r\bra{e_0}X_2\ket{e_2})&=0 \\
\Re(i \cosh r \bra{e_0}X_1\ket{e_1} - i \sinh r \bra{e_0}X_1\ket{e_2})&=0 \\
\Re(i \cosh r \bra{e_0}X_2\ket{e_1} -i \sinh r \bra{e_0}X_2\ket{e_2} )&=1\;. \label{eq_con6}
\end{align}
The matrix $Z_\theta$ in Eq.~(\ref{eq_zmat}) is given by
\begin{equation}
Z_\theta=
\begin{pmatrix}
\tr(\rho_0 X_1 X_1) & \tr(\rho_0 X_1 X_2) \\
\tr(\rho_0 X_2 X_1) & \tr(\rho_0 X_2 X_2)
\end{pmatrix}\;.
\end{equation}
Because we are interested in the minimization of Eq.\ (\ref{eq_hol}), we
can set to zero all components of $X_1$ and $X_2$ not involved in the
constraints Eq.\ (\ref{eq_con1}--\ref{eq_con6}) or not complex
conjugates of components involved in constraints. Define the
components in terms of their real and imaginary parts:
\begin{align}
\bra{e_0}X_1\ket{e_1}&=t_1+i j_1 \nonumber \\
\bra{e_0}X_1\ket{e_2}&=s_1+i k_1 \nonumber \\
\bra{e_0}X_2\ket{e_1}&=t_2+i j_2 \nonumber \\
\bra{e_0}X_2\ket{e_2}&=s_2+i k_2 \;.
\end{align}
And so 
\begin{equation}
Z_\theta=
\begin{tiny}
\begin{pmatrix}
t_1^2+j_1^2+s_1^2+k_1^2 & t_1t_2+j_1j_2+s_1s_2+k_1k_2+i(j_1t_2-j_2t_1+k_1s_2-k_2s_1) \\
t_1t_2+j_1j_2+s_1s_2+k_1k_2+i(j_2t_1-j_1t_2+k_2s_1-k_1s_2) & t_2^2+j_2^2+s_2^2+k_2^2
\end{pmatrix},
\end{tiny}
\end{equation}
The Holevo function from Eq.\ (\ref{eq_hol2}) becomes
\begin{align}
\label{eq_h1}
h=\underbrace{t_1^2+t_2^2+s_1^2+s_2^2+j_1^2+j_2^2+k_1^2+k_2^2}_{f}+2 \textrm{abs}\underbrace{\left\{j_1t_2-j_2t_1+k_1s_2-k_2s_1\right\}}_g.
\end{align}
Using the constraints Eq.\ (\ref{eq_con3}--\ref{eq_con6}),
we can eliminate four variables by making the substitutions
\begin{align}
t_1&= \sech r -s_1 \tanh r \nonumber \\
j_1&= k_1 \tanh r \nonumber \\
t_2&= - s_2 \tanh r \nonumber \\
j_2&=- \sech r +k_2 \tanh r\;.
\end{align}
The problem now is to minimize $h$ over the four remaining variables. This is accomplished in appendix \ref{sec_min}. We find that the minimum of $h$ and the Holevo CR bound
is $C_\text{H}=4\exp(-2r)$. This is equal to the sum of the MSE obtained
from the dual homodyne measurement. Hence, the dual homodyne measurement is the optimal measurement and the Holevo CR bound is tight.

\subsection{Conclusion}

We calculated the Holevo CR bound for a pure single-mode squeezed state probe experiencing a unknown displacement of $D(\theta)$. As expected, this equals the RLD CR bound which is known to be tight. 

We calculated the Holevo CR bound for a pure two-mode squeezed state probe to be $C_\text{H}=4\exp(-2r)$. This bound is superior to the SLD and RLD CR bound found by~\cite{Genoni2013,Gao2014}. The dual homodyne measurement obtains our bound indicating that the bound is tight.

Our calculation relied on the probe state being pure, so a natural extension to our work is to find the Holevo CR bound for $N>0$, i.e.\ when the probe is a two mode squeezed thermal state, and to determine whether the dual homodyne measurement is also optimal is this case.

\textbf{Acknowledgements} This research is supported by the
Australian Research Council (ARC) under the Centre of Excellence for
Quantum Computation and Communication Technology (CE110001027). We would also like to thank Jing Yan Haw for comments on the paper.

\appendix

\section{Calculations required for results}

\subsection{Derivatives of displacement operator}
\label{sec_dD}

To calculate the derivatives of the displacement operator, we use the Baker-Campbell-Hausdorff identity
\begin{equation}
\e^{A+B} = \e^A\e^B\e^{-\frac{1}{2}[A,B]},
\end{equation}
where $A$ and $B$ are operators that do not commute, but commute with $[A,B]$.
\begin{align}
\frac{\partial}{\partial\theta_1} D(\theta) &= \frac{\partial}{\partial\theta_1} \exp\left(\frac{i}{2} \theta_2 Q - \frac{i}{2} \theta_1 P \right) \\
&= \frac{\partial}{\partial\theta_1} \exp\left( - \frac{i}{2} \theta_1 P \right) \exp\left(\frac{i}{2} \theta_2 Q\right) \exp\left(-\frac{1}{8}\theta_1\theta_2[P,Q]\right) \\ 
&= \left(-\frac{i}{2}P -\frac{1}{8}\theta_2[P,Q]\right) \exp\left( - \frac{i}{2} \theta_1 P \right) \exp\left(\frac{i}{2} \theta_2 Q\right) \exp\left(-\frac{1}{8}\theta_1\theta_2[P,Q]\right) \\
&= \left(-\frac{i}{2}P -\frac{1}{8}\theta_2[P,Q]\right) D(\theta) \\
&=\left(-\frac{i}{2}P+\frac{i}{4}\theta_2\right) D(\theta),
\end{align}
where we have used $[Q,P]=2i$. Similarly,
\begin{align}
\frac{\partial}{\partial\theta_2} D(\theta) &= \frac{\partial}{\partial\theta_1} \exp\left(\frac{i}{2} \theta_2 Q - \frac{i}{2} \theta_1 P \right) \\
&= \frac{\partial}{\partial\theta_2} \exp\left(\frac{i}{2} \theta_2 Q\right) \exp\left( - \frac{i}{2} \theta_1 P \right) \exp\left(-\frac{1}{8}\theta_1\theta_2[Q,P]\right) \\ 
&= \left(\frac{i}{2}Q -\frac{1}{8}\theta_1[Q,P]\right) \exp\left(\frac{i}{2} \theta_2 Q\right) \exp\left( - \frac{i}{2} \theta_1 P \right) \exp\left(-\frac{1}{8}\theta_1\theta_2[Q,P]\right) \\
&= \left(\frac{i}{2}Q -\frac{1}{8}\theta_1[Q,P]\right) D(\theta) \\
&=\left(\frac{i}{2}Q-\frac{i}{4}\theta_1\right) D(\theta).
\end{align}

\subsection{Performing the minimization for the two-mode probe}
\label{sec_min}

In order to contend with the absolute value in Eq.\ (\ref{eq_h1}), we
consider two cases, case 1: $g$ is greater or equal to zero, and
case 2: $g$ is less than zero. We consider the case when $r\neq
0$. When $r=0$, the problem reduces to a single-mode probe and is
discussed in section \ref{sec_singlemode}.

\subsection*{case 1: $g\geq 0$}
Minimize
\begin{align}
\label{eq_hcase1}
h=f+2g
\end{align}
subject to
\begin{align}
\label{eq_gcase1}
g \geq 0\;.
\end{align}
From the Karush-Kuhn-Tucker conditions, necessary (but not sufficient) conditions for the minimum are 
\begin{align}
-\grad{(f+2g)}&=-\lambda\grad{g} \label{eq_kkt1} \\
g&\geq 0 \label{eq_kkt2} \\
\lambda&\ge 0 \label{eq_kkt3}  \\
\lambda g&=0 \label{eq_kkt4} 
\end{align}
where
$\grad=(\frac{\partial}{\partial s_1},\frac{\partial}{\partial
  k_2},\frac{\partial}{\partial k_1},\frac{\partial}{\partial
  s_2})$. Equation (\ref{eq_kkt1}) becomes
\begin{align}
\begin{pmatrix}
-2\cosh2r & \lambda-2&0&0\\
\lambda-2&-2 \cosh 2r&0&0\\
0&0&2 \cosh 2r& \lambda-2\\
0&0& \lambda-2&2 \cosh 2r
\end{pmatrix}
\begin{pmatrix}
s_1\\
k_2\\
k_1\\
s_2
\end{pmatrix}=
\begin{pmatrix}
-\lambda \sinh r\\
-\lambda \sinh r\\
0\\
0
\end{pmatrix}\;.
\end{align}
For $r\neq 0$, this set of equations has no solutions when $\lambda=4
\cosh^2 r$. We thus
consider $\lambda\neq 4 \cosh^2 r$, and find
\begin{align}
\label{eqco}
s_1=k_2 &= \frac{\lambda \sinh r}{4\cosh^2 r-\lambda}\;\text{ and }\; k_1=s_2 = 0\;.
\end{align}
From Eq.\ (\ref{eq_kkt4}) we have that either $\lambda=0$ or $g=0$. Let us consider each case seperately. 

\subsection*{case 1a: $\lambda=0$}
When $\lambda=0$ the solutions~(\ref{eqco}) become
\begin{align}
s_1=k_2 = k_1=s_2 = 0\;,
\end{align}
which gives $g=-\sech^2 r<0$ and violates condition (\ref{eq_kkt2}).
Hence this is not a valid solution.

\subsection*{case 1b: $g=0$}
When $g=0$, we solve for $\lambda$ to get $\lambda=4 \e^{\pm
  r} \cosh r $. Both are valid solutions and give $h=4 \e^{\pm 2r}$. Although $h=4 \e^{2r}$ satisfies the Karush-Kuhn-Tucker conditions, is not the minimum so we can ignore this solution.

\subsection*{case 2: $g<0$}
Minimize
\begin{align}
\label{eq_hcase1}
h=f-2g
\end{align}
subject to
\begin{align}
\label{eq_gcase1}
g < 0
\end{align}
The Karush-Kuhn-Tucker conditions now become
\begin{align}
-\grad{(f-2g)}&= \lambda\grad{g} \label{eq_kkt12} \\
g& \leq 0 \label{eq_kkt22} \\
\lambda&\ge 0 \label{eq_kkt32}  \\
\lambda g&=0 \label{eq_kkt42} \;.
\end{align}
Since we require $g<0$, condition~(\ref{eq_kkt42}) implies $\lambda=0$
for which condition~(\ref{eq_kkt12}) becomes 
\begin{align}
\begin{pmatrix}
-\cosh2r & 1&0&0\\
1& - \cosh 2r&0&0\\
0&0& \cosh 2r& 1\\
0&0& 1& \cosh 2r
\end{pmatrix}
\begin{pmatrix}
s_1\\
k_2\\
k_1\\
s_2
\end{pmatrix}=
\begin{pmatrix}
-2 \sinh r\\
-2 \sinh r\\
0\\
0
\end{pmatrix}\;.
\end{align}
For $r\neq 0$, this has the solution $s_2=k_1=0$, $s_1=k_2=\csch r$
which gives $g=\csch^2 r >0$, hence is not a valid solution. 

\subsubsection*{solution}
Putting it all together, the smallest solution satisfying the Karush-Kuhn-Tucker conditions, and hence the minimum of $h$ and the Holevo bound
is $C_\text{H}=4\exp(-2r)$. 

\bibliographystyle{unsrt}

\begin{thebibliography}{10}

\bibitem{Holevo1976a}
A.~S. Holevo.
\newblock Noncommutative analogues of the cram\'er-{Rao} inequality in the
  quantum measurement theory.
\newblock {\em Lecture Notes in Mathematics}, pages 194--222, 1976.

\bibitem{Holevo2011}
Alexander~S Holevo.
\newblock {\em Probabilistic and statistical aspects of quantum theory},
  volume~1.
\newblock Springer Science \& Business Media, 2011.

\bibitem{Genoni2013}
MG~Genoni, MGA Paris, G~Adesso, H~Nha, PL~Knight, and MS~Kim.
\newblock Optimal estimation of joint parameters in phase space.
\newblock {\em Physical Review A}, 87(1):012107, 2013.

\bibitem{Gao2014}
Yang Gao and Hwang Lee.
\newblock Bounds on quantum multiple-parameter estimation with gaussian state.
\newblock {\em The European Physical Journal D}, 68(11):1--7, 2014.

\bibitem{Helstrom1967}
CW~Helstrom.
\newblock Minimum mean-squared error of estimates in quantum statistics.
\newblock {\em Physics letters A}, 25(2):101--102, 1967.

\bibitem{Helstrom1969}
Carl~W Helstrom.
\newblock Quantum detection and estimation theory.
\newblock {\em Journal of Statistical Physics}, 1(2):231--252, 1969.

\bibitem{Yuen1973}
H~Yuen and Melvin Lax.
\newblock Multiple-parameter quantum estimation and measurement of
  nonselfadjoint observables.
\newblock {\em IEEE Transactions on Information Theory}, 19(6):740--750, 1973.

\bibitem{Paris2009}
Matteo~GA Paris.
\newblock Quantum estimation for quantum technology.
\newblock {\em International Journal of Quantum Information},
  7(supp01):125--137, 2009.

\bibitem{Petz2011}
D.~Petz and C.~Ghinea.
\newblock {\em Introduction to Quantum Fisher Information}, chapter~15, pages
  261--281.
\newblock World Scientific, Jan 2011.

\bibitem{Hayashi2006}
Masahito Hayashi and Keiji Matsumoto
\newblock Asymptotic performance of optimal state estimation in quantum two level system.
\newblock {\em arXiv:quant-ph/0411073}, 2006.

\bibitem{Yamagata2013}
Koichi Yamagata, Akio Fujiwara, and Richard~D. Gill.
\newblock Quantum local asymptotic normality based on a new quantum likelihood
  ratio.
\newblock {\em The Annals of Statistics}, 41(4):2197--2217, Aug 2013.

\bibitem{Matsumoto2002}
K~Matsumoto.
\newblock A new approach to the {C}ramer-{Rao}-type bound of the pure-state
  model.
\newblock {\em Journal of Physics A: Mathematical and General},
  35(13):3111--3123, Mar 2002.

\bibitem{Fujiwara1994a}
Akio Fujiwara.
\newblock Linear random measurements of two non-commuting observables.
\newblock {\em Math. Eng. Tech. Rep}, 94(10), 1994.

\bibitem{Fujiwara1999}
Akio Fujiwara and Hiroshi Nagaoka.
\newblock An estimation theoretical characterization of coherent states.
\newblock {\em Journal of Mathematical Physics}, 40(9):4227--4239, 1999.

\bibitem{Fujiwara1994}
Akio Fujiwara.
\newblock Multi-parameter pure state estimation based on the right logarithmic
  derivative.
\newblock {\em Math. Eng. Tech. Rep}, 94(9):94--10, 1994.

\end{thebibliography}

\end{document}